# Hydrogen clustering in bcc metals: atomic origin and strong stress anisotropy

Jie Hou[1], Xiang-Shan Kong[2,3], C.S. Liu[3], Jun Song[1*]

[1] Department of Mining and Materials Engineering, McGill University, Montreal, Quebec H3A 0C5, Canada.

[2] Key Laboratory for Liquid-Solid Structural Evolution and Processing of Materials, Ministry of Education, Shandong University, Jinan, Shandong 250061, PR China

[3] Key Laboratory of Materials Physics, Institute of Solid State Physics, Chinese Academy of Sciences, PO Box 1129, Hefei 230031, China

**Abstract**

Hydrogen (H) induced damage in metals has been a long-standing woe for many industrial applications. One form of such damage is linked to H clustering, for which the atomic origin remains contended, particularly for non-hydride forming metals. In this work, we systematically studied H clustering behavior in bcc metals represented by W, Fe, Mo, and Cr, combining first-principles calculations, atomistic and Monte Carlo simulations. H clustering has been shown to be energetically favorable, and can be strongly facilitated by anisotropic stress field, dominated by the tensile component along one of the <001> crystalline directions. We showed that the stress effect can be well predicted by the continuum model based on H formation volume tensor, and that H clustering is thermodynamically possible at edge dislocations, evidenced by nanohydride formation at rather low levels of H concentration. Moreover, anisotropy in the stress effect is well reflected in nanohydride morphology around dislocations, with nanohydride growth occurring in the form of thin platelet structures that maximize one <001> tension. In particular, the <001> type edge dislocation, with the <001> tensile component maximized, has been shown to be highly effective in facilitating H aggregation, thus expected to play an important role in H clustering in bcc metals, in close agreement with recent experimental observations. This work explicitly and quantitatively clarifies the anisotropic nature of stress effect on H energetics and H clustering behaviors, offering mechanistic insights critical towards understanding H-induced damages in metals.

1. Introduction

Since its first discovery in 1875 [1], hydrogen (H) induced damage in metals has been recognized as a central and long-standing problem in materials science. H may reduce ductility and promote cracking in metals [2-4] to cause premature damages of load-bearing components, which potentially leads to catastrophic failures. H is also known to induce blisters or bubbles in metals to greatly undermine their structural integrity and safety [5-8]. Consequently, it poses great threats to various industrial applications involving metallic materials. Aiming to understand and even mitigate H induced damages, decades of efforts were therefore made to study the behavior of H in metals. Numerous advanced experimental techniques, such as *in situ* environmental transmission electron microscopy [5, 9] or atom probe tomography [10, 11], were employed to reveal nanoscale details of H damages. Meanwhile, different mechanisms and theories, including H enhanced decohesion [12], H enhanced local plasticity [13-15], H enhanced vacancy formation [16, 17], among others, were proposed to explain different H induced damage phenomena.

---

[*] Author to whom correspondence should be addressed. Email: jun.song2@mcgill.ca

Yet despite these substantial research efforts and great progresses made, there remain numerous critical gaps in our understanding of H damage processes and mechanisms. One particular aspect that remains highly debated is H clustering in metals, since most H damaging mechanisms proposed involve H concentrated in a local region [18, 19]. H clustering starts from the simply picture of interaction between two H interstitials in metal, which are either repulsive or weakly attracted to each other [20-25]. It was commonly believed that in non-hydride forming metals like tungsten (W), iron (Fe), and Nickel (Ni), H damages cannot initiate from H clustering, and pre-existing defects (e.g., vacancies [26, 27], grain boundaries [28, 29], *etc.*) are necessary for H damages to occur. This perspective, however, contradict with extensive experimental evidences where H was shown to induce damages even in metals with very low defect density. A typical example is H induced bubbling/blistering in well annealed pure W with coarse-grained or single crystal structure [30-35], where influence of pre-existing defects is minimized, making H clustering a plausible mechanism. To fundamentally understand such discrepancy, there have been a few atomistic studies that investigate the interaction between multiple H, and H clustering in non-hydride forming metals [35, 36]. For instance, recent first-principles calculations have shown that H in tungsten (W) can form platelet-like self-clusters along {001} planes, with binding energy between H gradually increases as the self-cluster grows in size [35, 36]. Further to those first-principles studies, large-scale atomistic simulations have also been conducted to examine H self-trapping in W [37] employing an interatomic potential fitted to reproduce H-H interactions from first-principles database [38], demonstrating formation of platelet-like H self-clusters, induced by dislocations or homogeneous stresses, or at high H concentration [37]. Similar results were also found in some other metals with H clusters observed in stressed regions [4, 22, 39, 40]. Nonetheless, many of these studies focused on H-defect interaction, while limited attention was directed towards H cluster itself.

These recent findings clearly highlight the feasibility of H clustering in metals, yet many atomic details remain unclear or controversial. One basic but crucial question is how stress influences the clustering of H. Despite numerous experiments [41-44] and simulations [4, 22, 39, 40] have suggested that H clustering can be facilitated by stress, the clustering behavior shows notable distinction in different metals. In general, H clustering in face centered cubic (fcc) metals such as Ni is mainly related to isotropic stresses [13, 30, 31], and can be well described by isotropic linear elastic theory [13, 30]. While in body centered cubic (bcc) metals, including W, Fe, molybdenum (Mo), and chromium (Cr), both individual H dissolution [45] and H clustering [4, 37] can be highly anisotropic, with recent simulations specifically showing that isotropic stress along cannot induce H clustering in W [37]. Moreover, a recent experimental work [7] showed that <001> type edge dislocations (rather than the commonly discussed <111>/2 type ones) play a key role in inducing H blisters in W, and proposed a blister nucleation mechanism based on H clustering around the <001> edge dislocation. The above hints the necessity of more dedicated studies on the anisotropic effect. Unfortunately, most previous studies generally focus on individual H in metals [45-47], or simply adopt the isotropic approximation [4, 22, 39, 48, 49] that can be problematic to H clusters in bcc metals. Quantitative explanation and evaluation of anisotropic stress effects on H clustering in bcc metals thus remain unavailable, rendering a critical knowledge deficit in understanding H damages related to H clustering in metallic materials.

In the present work, we systematically investigated H clustering behavior in non-hydride-forming bcc metals represented by W, Fe, Mo, and Cr, combining first-principles calculations, atomistic and Monte Carlo simulations. The formation energetics of different H structures, with or without the influence of external stresses, have been examined, with the effect of stress found to be highly anisotropic. The underlying chemomechanical origin of such anisotropic effect was clarified, enabling accurate continuum predictions of formation enthalpies for H within different structures. Large-scale simulations were then performed to directly study the morphology and extend of H clustering, assisted by stress fields from dislocations, under different H chemical potentials. In the end, the implications of our results to interpreting experimental observations and general understanding of H-induced damages in metals were discussed, and key findings were summarized.

## 2. Computational methods

Different computational methods are integrated in our study to provide a systematic investigation of H clustering behavior. First, individual H interstitials and ordered H structures were examined using density functional theory (DFT) calculations to provide accurate information on H formation enthalpies and volume tensors. Then molecular dynamics (MD) simulations based on empirical interatomic potentials were employed to expand the investigation to more general nanosized H clusters. MD was further combined with grand canonical Monte Carlo (GCMC) to enable hybrid MD+GCMC simulations to examine thermodynamics of H cluster formation in bulk lattice and at dislocations. Below the computational details are elaborated.

### 2.1 Density functional theory (DFT) calculations

All DFT calculations in this work were performed using the Vienna *ab initio* simulation package (VASP) [50, 51] with Blochl's projector augmented wave (PAW) potential method [52]. All the outer *d*-shell and *s*-shell electrons of metals were treated as valence electrons (1*s* for H, 3*d* and 4*s* for Cr and Fe, 4$d$ and 5$s$ for Mo, 5$d$ and 6$s$ and W). Effects of spin polarization were considered in calculations for Cr and Fe systems. The exchange-correlation energy functional was described with the generalized gradient approximation (GGA) as parameterized by Perdew–Wang (PW91) [53]. A 500 eV plane wave cutoff was adopted with convergence criteria for energy and atomic force were set as $10^{-6}$ eV and 0.01 eV/Å respectively. The partial occupancies were determined using first order Methfessel-Paxton method [54] with a smearing width of 0.2 eV. Relaxations of atomic configuration and adjustment of the shape and size of the super-cell were performed until the system converges to designated stress state within a criterion of 0.01 GPa [55]. Super-cell sizes and k-point grids for different calculations are summarized in Table 1, and additional calculations have been performed to ensure accuracy and no-size dependence of the results (see Appendix A for details). Zero-point energy correction is included for H atoms by summing up their ground state vibration energies to ensure accurate description of the energetics.

Table 1. Super-cell sizes and k-points (for DFT calculations) used in simulating different H-containing systems. $a_0$ is the lattice parameter of the relate bcc metal.

|  | Super-cell | k-points | Method |
| --- | --- | --- | --- |
| Individual H interstitial | $4 \times 4 \times 4\ a_0^3$ | $3 \times 3 \times 3$ | DFT/EAM |
| Infinite hydride | $4 \times 4 \times 4\ a_0^3$ | $3 \times 3 \times 3$ | DFT/EAM |
| Infinite H plane | $3 \times 3 \times 12\ a_0^3$ | $4 \times 4 \times 1$ | DFT/EAM |
| Nanohydride | $50 \times 50 \times 50\ a_0^3$ | N/A | EAM |
| Dislocation-H | $25 \times 25 \times 1.4\sim1.9\ nm^3$ | N/A | EAM |

**2.2 Empirical potential based simulations**

Simulations based on empirical potentials were carried out using the Large-scale Atomic/Molecular Massively Parallel Simulator (LAMMPS) package [56] with the results visualized using the OVITO software [57]. These calculations were limited to W-H and Fe-H systems due to the availability of interatomic potentials. The embedded atom method (EAM) potential developed by Wang et al. [38] and EAM potential developed by Ramasubramaniam et al. [58] (with modification on H-H attraction according to Ref. [4]) were adopted for the W-H and Fe-H systems respectively. These two potentials were chosen as they were fitted to reproduce H-H interaction (besides other essential material properties) from DFT database. Similar to that in the DFT calculations, relaxations of atomic configuration and adjustment of the shape and size of the super-cell were performed until the system converges to designated stress state. The simulation cell sizes used are also summarized in Table 1.

To study the influence of dislocation on H clustering, we investigated (100)[001] and $(01\bar{1})[111]/2$ type edge dislocations in this work. Orthogonal simulation cells with basis vectors along $[001] \perp [100] \perp [010]$ and $[111] \perp [01\bar{1}] \perp [2\bar{1}\bar{1}]$ were used for two dislocations respectively, with free surface condition applied along the second direction (slip plane normal) and periodic boundary condition applied on the other two directions (dislocation line and burgers vector directions). The size of simulation cells is around $25 \times 25 \times 1.4\sim1.9\ nm^3$. Each dislocation model was relaxed via energy minimization [59], following which hybrid MD+GCMC simulations were performed to obtain H distribution around the dislocation. In the GCMC part, H atoms were randomly inserted/deleted according to the GCMC algorithm [60] to achieve a designated H chemical potential within a constant volume (i.e., μVT). In the MD part, isobaric-isothermal ensemble (i.e., NPT with P=0) with Nosé–Hoover thermostat was adopted and a timestep of 1 fs was used. Overall, the MD and GCMC steps were performed alternatively and iteratively with equal proportions, i.e., 100 MD timesteps after 100 GCMC attempts of H insertion/deletion. All MD and GCMC are performed with temperature T=300 K. The simulations were terminated when the system is fully converged with no further change in H number/structure, or when a total of 40,000,000 steps (20,000,000 MD steps + 20,000,000 GCMC steps) are reached.

### 2.3 Formation enthalpy and volume tensor calculation

In DFT and MD calculations, under a given stress tensor $\tilde{\sigma}^{ext}$, the H formation enthalpy $H_f^H(\tilde{\sigma}^{ext})$ for a H-containing system is calculated as:

$$H_f^H(\tilde{\sigma}^{ext}) = \frac{E_{bulk}^H - E_{bulk}}{N} - \frac{1}{2}E^{H_2} - \tilde{\sigma}^{ext} : \tilde{\Omega}^H(\tilde{\sigma}^{ext}), \qquad (1)$$

where $N$ is the number of H atoms, $E_{bulk}^H$ and $E_{bulk}$ are energies of the system with and without H under the given stress, $E^{H_2}$ is the energy of a $H_2$ molecule in vacuum. $\tilde{\Omega}^H(\tilde{\sigma}^{ext})$ is the formation volume tensor of H under the given stress, and is calculated by [61]:

$$\Omega_{ii}^H = \frac{1}{N}V_{bulk}\varepsilon_{ii}^H\left[1 + \frac{1}{2}\left(\varepsilon_{jj}^H + \varepsilon_{kk}^H\right) + \frac{1}{3}\varepsilon_{jj}^H\varepsilon_{kk}^H\right], \qquad (2)$$

where $V_{bulk}$ is the system volume without H, and $\varepsilon_{ii}^H$ is the tensor component of H induced strain to the system (all shear components are 0 in this work due to the tetragonal symmetry of related structures). The formation volume tensor defined in this way effectively quantifies the H induced volume change in each direction. Note for small strains, all high order terms on the right side can be neglected, which gives the commonly used infinitesimal form, $\Omega_{ij}^H = V^{bulk}\varepsilon_{ij}^H/N$.

### 3. Results and Discussion

### 3.1 H in different structures under stress free condition

We start off with the typical interstitial sites in bcc metals, i.e., the tetrahedral interstitial (TIS) and octahedral interstitial (OIS) sites individual H occupies, as illustrated in Figs. 1a-b. A closer look at the local symmetry associated with TIS and OIS sites, we note that OIS, with two neighboring metal atoms at a much closer distance than the other four, inherently possesses a tetragonal symmetry with four-fold axis along one of the <001> directions. Therefore, OIS is expected to show different responses to stresses along and perpendicular to the symmetry axis. Similar tetragonal symmetry can also be found for the TIS. Considering such symmetry, in the following, we will focus our discussion on structures with the four-fold axis along [001] direction (i.e., structures shown in Fig. 1).

Next we consider a few representative structures where H atoms aggregate in an ordered manner, to put our discussion in the context of H clustering, illustrated in Figs. 1c-f. In particular, Fig. 1c shows a planar H cluster comprised by TIS H arranged along a (001) plane (denoted as $HC_T^{001}$ below), which was shown to be more stable than individual TIS H by previous DFT studies [35, 36]. A similar planar cluster is also constructed by arranging OIS H along (001) plane ($HC_O^{001}$ for short) and is shown in Fig. 1d. Along with these two-dimensional clusters, we further considered hydride structures that can be viewed as three-dimensional extensions of these clusters, shown in Figs. 1e and 1f, being di-hydride structure with H filling all [001] type TIS ($HD_T^{MH2}$ for short) and rock-salt mono-hydride structure with H filling all [001] type OIS ($HD_O^{MH}$ for short) respectively. The two hydrides are also the most commonly seen structures for mono- and di-hydrides for transition metals [62]. Note that both $HD_T^{MH2}$ and $HD_O^{MH}$ structures are heavily distorted tetragonally and effectively become fcc lattices (or bct lattice with c/a= $\sqrt{2}$) via reverse Bain transformation. Such H induced bcc to fcc transformation was also observed in previous MD simulations for bcc metals, e.g., Fe [4] and W [37].

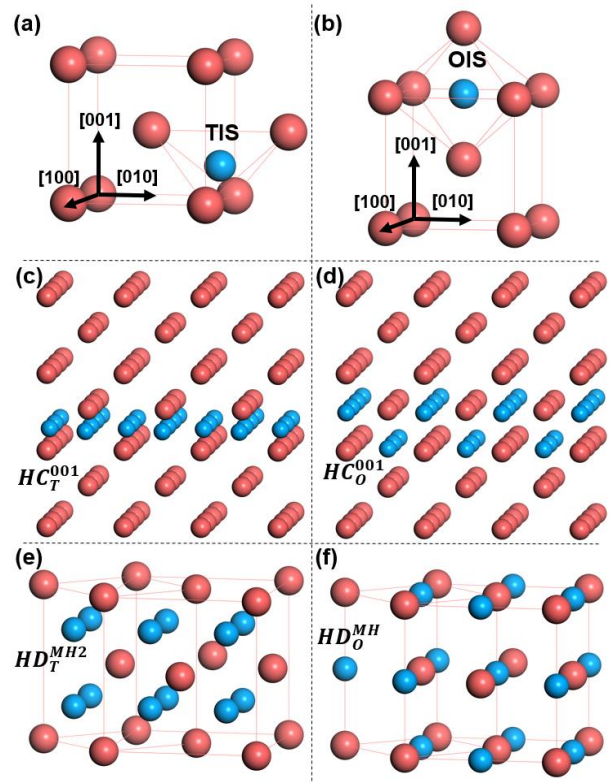

Figure 1. Schematics of different structures of H (blue atoms) in a bcc lattice (metal atoms colored red). (a) Interstitial H at TIS, (b) interstitial H at OIS, (c) planar structure ($HC_T^{001}$) comprised by TIS H arranged along (001) plane, (d) planar structure ($HC_O^{001}$) comprised by OIS H arranged along (001) plane, (e) di-hydride ($HD_T^{MH2}$) with fluorite structure, (f) mono-hydride ($HD_O^{MH2}$) with rock-salt structure.

DFT calculations were then performed to examine the formation enthalpies for those structures constructed. Fig. 2 shows the values of formation enthalpy under zero stress condition, where all structures were fully relaxed without constrain. A few observations can be made from these results. For individual H interstitial, TIS is always more preferable than OIS, in agreement with previous studies [45, 63-65]. Across the four bcc metals, there is a general rising trend in the formation enthalpy in the order of Fe, Cr, Mo and W, indicative of their different susceptibility of H infiltration and clustering. The lowest point in the formation enthalpy is found for $HD_O^{MH}$ in Fe, which is positive and consistent with the general notion of those metals being non-hydride forming systems. We should note that fcc hydrides were experimentally observed in austenitic steels under electrolytic charging [66-68], with lattice parameters around 3.72~3.76 Å that agree well with our DFT-calculated value of 3.76 Å for $HD_O^{MH}$ in Fe. These experimentally observed hydrides are however unstable and decompose when the charging is stopped or at room temperature [66-68], confirming with the small positive formation enthalpy of $HD_O^{MH}$ in Fe. Yet meanwhile we can also note that H is always energetically favorable in clustered structures than individual H interstitials, suggesting the possibility of H clustering.

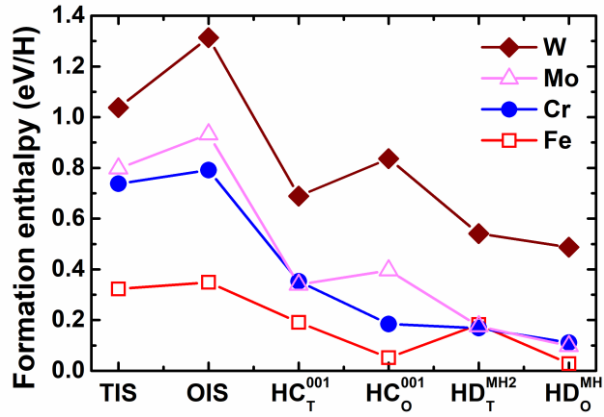

Figure 2. Formation enthalpies for H in different structures in bcc metals under stress-free condition.

## 3.2 Effect of external stress on formation enthalpy of H

As discussed above, H clustering in bcc metals is energetically favorable and expected to show anisotropic responses to stresses due to the tetragonal symmetry. To quantify such anisotropic behavior, we performed calculations for different H structures with uniaxial stresses applied respectively along [001] and [100] directions ([010] is equivalent to [100] due to the tetragonal symmetry, thus not discussed in the following). Fig. 3 present the calculated formation enthalpies of H (denoted by symbols), showing similar patterns across different bcc metals while notable distinctions for H with different structures. Therefore, we first limit our focus on different H structures using W-H system as a representative (i.e., Figs. 3a-c). Starting with H interstitials in W, as shown in Fig. 3a, it's evident that the formation enthalpy for OIS H is most sensitive to the [001] stress, but much less to [100] stresses. Meanwhile for the TIS H, the formation enthalpy shows similar responses to stresses along different directions, indicating a nearly isotropic response to stresses. It is also worth noting OIS H show a much larger drop in its formation enthalpy under uniaxial tension along [001], rendering it even energetically more favorable than TIS H for large tensile stresses. This is also reflected in the DFT data for TIS H at high tensile stresses (cf. Fig. 3a), which are observed to deviate from the isotropic pattern. Such deviation is caused by some TIS H (particularly those neighboring OIS) becoming unstable and shifting towards the OIS site. Similar TIS-OIS transition under uniaxial tension has also been reported by other studies [46]. In addition, we also notice OIS H become slightly more stable under compressive stresses along [100] (and the equivalent [010]) direction, reflecting a TIS-OIS transition under other stress states, e.g., transition induced by large biaxial compressive strain that reported in Ref. [45].

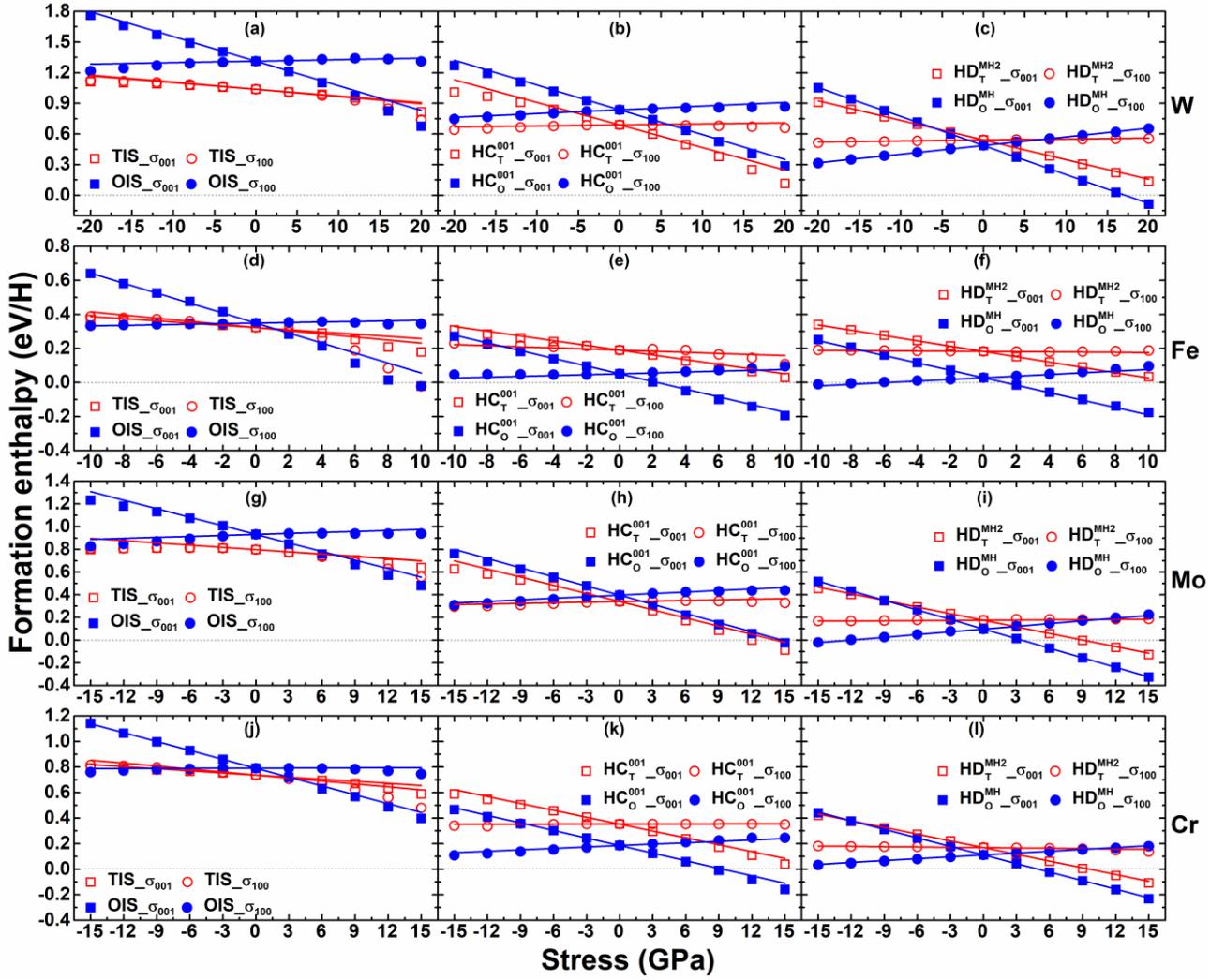

Figure 3. Formation enthalpies for (a) H at TIS and OIS, (b) planar H clusters, and (c) hydrides under uniaxial stresses along [001] and [100] directions (denoted by $\sigma_{001}$ and $\sigma_{100}$ respectively) in W. (d-f), (g-i), and (j-l) respectively show similar results for Fe, Mo, and Cr. Symbols are data calculated from DFT, while lines are corresponding linear elasticity predictions from Eq. 3. Note that all the structures were constructed with the four-fold axis along [001] direction as demonstrated in Fig. 1.

Further to the interstitial H, we calculated stress-dependent formation enthalpies of H in aggregated structures in W, with results shown in Fig. 3b and Fig. 3c. Here, we see that all clustered H structures show anisotropic response to stresses along different orientations. For the $HC_T^{001}$ and $HC_O^{001}$ planar H clusters (cf. Fig. 3b), their formation enthalpies exhibit very similar responses as the case of OIS H, with strong sensitivity to the [001] stress, but rather insensitive to [100] stresses. Likewise, the formation enthalpies of the two hydrides, i.e., $HD_T^{MH2}$ and $HD_O^{MH}$ (cf. Fig. 3c), also exhibit highly anisotropic responses to stresses, again with their formation enthalpies decrease considerably under tension along [001]. On the other hand, they show different behaviors in response to the [100] stress. The formation enthalpy of $HD_T^{MH2}$ di-hydride is hardly affected by the [100] stress, while that of the $HD_O^{MH}$ mono-hydride undergoes notable decrease under compression along [100]. Furthermore, we can see that, accounting for influence from stresses along different directions, overall the $HD_O^{MH}$ mono-hydride is energetically more favorable than the $HD_T^{MH2}$ di-hydride, which explains why H tends to occupy octahedral sites in nanohydrides, as reported in previous MD studies [4, 37].

Apart from the W-H system, similar H formation enthalpy data for Fe, Mo, Cr is shown in Fig. 3d-l. We noticed that general trends of these data are almost identical across four bcc metals, with TIS H showing nearly isotropic response to stresses along different directions, while OIS H and clustered H demonstrate evident anisotropy. The similarity across different metals indicates that the anisotropic behavior is not an *ad hoc* feature but a universal property of H in bcc metals. To gain a more general insight into the anisotropic stress effect, Fig. 4 summarizes the results shown in Fig. 3 with all data consolidated into the three categories of interstitial, planar cluster and hydride. Note that for each category, only the minimum formation enthalpy value, accounting for loading along different directions, is shown here. For all four bcc metals, with the formation enthalpy ranked from low to high, we have the hydride form ($HD_O^{MH}$ is most stable for all metals), planar H cluster ($HC_T^{001}$ for W and Mo, $HC_O^{001}$ for Cr and Fe), and interstitial H. This further confirms that energetically H clustering would be feasible in these metals. One interesting observation from Fig. 4 is that both planar H and hydride structures can be further stabilized by stresses under both tension and compression, which suggests that H clustering may be facilitated by stresses, either externally applied or arising from microstructural entities such as crack [4], dislocation [37], or inclusion [69].

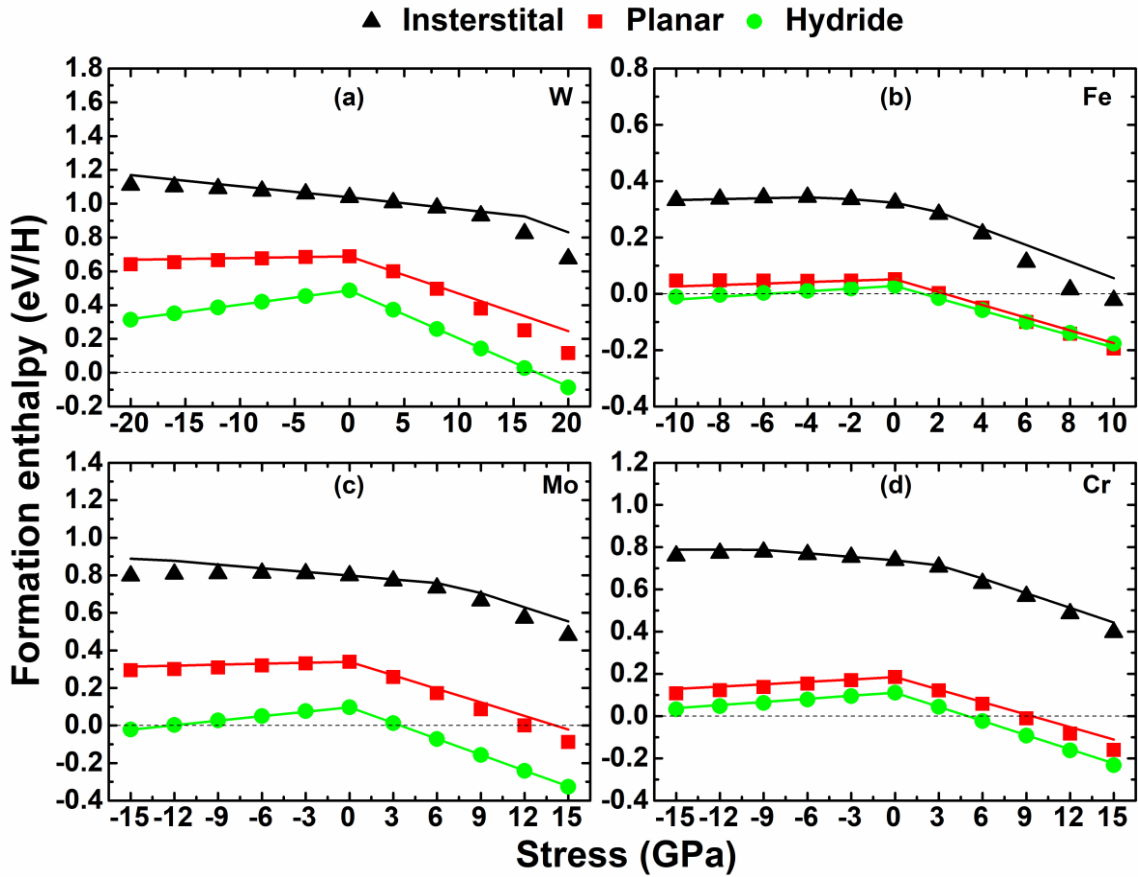

Figure 4. The minimum formation enthalpy for H at interstitial sites, in planar and hydride structures under uniaxial stresses along [001] and [100] directions (created based on the data shown

in Fig. 3) in (a) W, (b) Fe, (c) Mo, and (d) Cr. Symbols are data calculated from DFT, while lines are corresponding linear elasticity predictions based on Eq. 3.

## 3.3 Chemomechanical origin of anisotropic stress effect on H clustering

The above DFT calculations clearly show that the influence of stress on H clustering is highly anisotropic, which apparently relates to the tetragonal symmetry shown in Fig. 1. To reveal the atomic origin of such anisotropy, the structural tetragonality is quantified in terms of the formation volume tensor components of H, i.e., $\Omega_{ij}^H$, indicative of the volume changes induced by H along different directions, as previously described in Eq. 2. Figs. 5a-d present the different components of the formation volume tensor $\widetilde{\Omega}^H(0)$ for H in different structures under stress-free condition. Note that for the sake of clarity, below tensor components such as $\Omega_{xx}^H/\Omega_{yy}^H/\Omega_{zz}^H$ are denoted as $\Omega_{100}^H/\Omega_{010}^H/\Omega_{001}^H$ unless specified otherwise. From Figs. 5a-d we can see that $\widetilde{\Omega}^H(0)$ is generally highly anisotropic, showing significant tetragonal distortion with large $\Omega_{001}^H$ while small or negative $\Omega_{100}^H$. Nonetheless there is one exception of the case of H at TIS which exhibits nearly isotropic expansion (i.e., $\Omega_{001}^H \approx \Omega_{100}^H$). Moreover, we found that $\widetilde{\Omega}^H(0)$ is closely related to the stress-dependent H formation enthalpy that shown in Fig. 3. In particular, we note that $\widetilde{\Omega}^H(0)$ is directly correlated with the sensitivity of the formation enthalpy to stress. Such correlation is further illustrated in Fig. 5, where side-to-side comparison is made between the $\widetilde{\Omega}^H(0)$ components, i.e., $\Omega_{001}^H$ and $\Omega_{100}^H$, and the slopes of corresponding H formation enthalpy vs. stress plots from Fig. 3, showing an excellent correspondence.

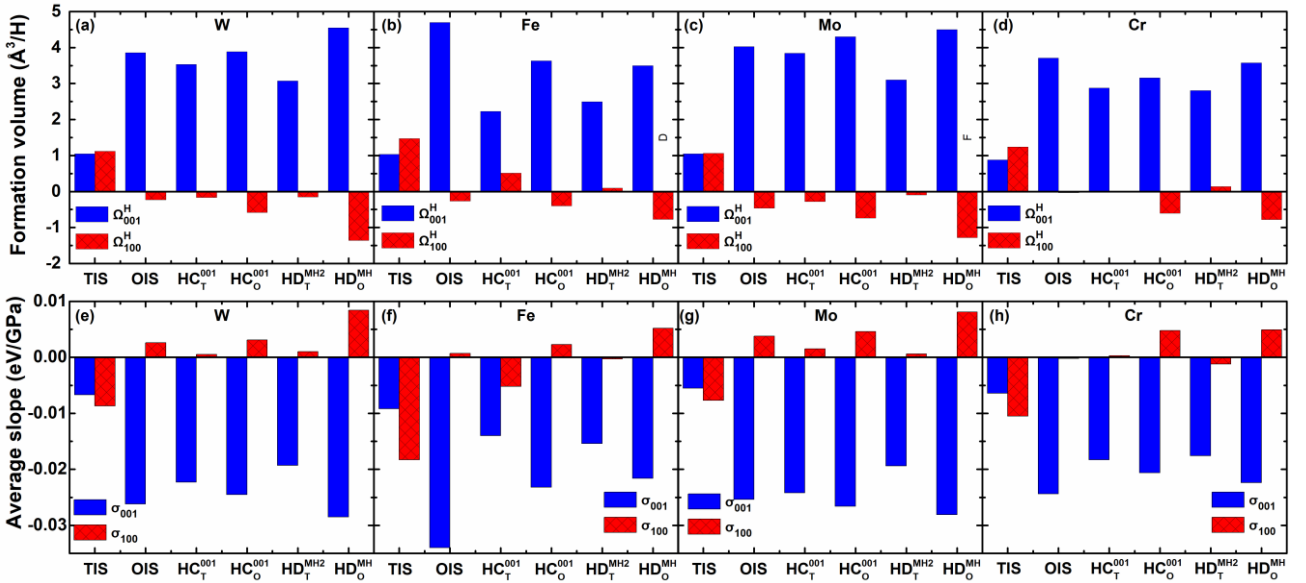

Figure 5. (a-d) Different components of formation volume tensor $\widetilde{\Omega}^H(0)$ for H in different structures in different bcc metals under stress-free condition. (e-h) Average slopes of H formation enthalpy change against stresses along [001] and [100] directions (denoted by $\sigma_{001}$ and $\sigma_{100}$

respectively), for H in different structures in different bcc metals, calculated by linear fittings of the DFT data shown in Fig. 3.

Fig. 5 clearly demonstrated $\widetilde{\Omega}^H(0)$ as a significant factor in determining the anisotropic influence of stress on H clustering. To understand the role played by $\widetilde{\Omega}^H(0)$, we note that, from continuum mechanics, the change in H formation enthalpy under external stress can be determined as the following, assuming linear elasticity:

$$H_f^H(\widetilde{\sigma}^{ext}) = H_f^H(0) - \widetilde{\sigma}^{ext}:\widetilde{\Omega}^H(0), \qquad (3)$$

where $H_f^H(0)$ represent the H formation enthalpy under zero stress (cf. Fig. 2). Note that Eq. 3 is equivalent to the commonly used form based on force-dipole tensor [70, 71], but is more convenient and straightforward in evaluating anisotropy (for comparison of the two methods, see Appendix B). Predictions from Eq. 3 are plotted alongside with the DFT data in Figs. 3-4, showing excellent agreement.

On base of Eq. 3, quantitative interpretation of the DFT results can then be made. Under uniaxial loading that shown in Fig. 3, the change in formation enthalpy depends only on the $\Omega_{ij}^H$ component along the stress direction. For instance, we demonstrated in Fig. 3 that OIS H and all H clusters show particular favor to the [001] tension, but being generally insensitive to the [100] stress. This directly derives from the fact that they have large $\Omega_{001}^H$ while small $\Omega_{100}^H$ (cf. Fig. 5a-5d). Meanwhile, the similar $\Omega_{001}^H$ and $\Omega_{100}^H$ components for TIS H also explain its nearly isotropic responses to stresses along different directions. In addition, with OIS H having much large $\Omega_{001}^H$ than TIS H, the prediction also captures the spontaneous TIS-OIS transition under large [001] tension that demonstrated in Fig. 3. In general, the anisotropic stress effects shown in Figs. 3-4 directly originate from different H formation volume along different directions, and can be quantitatively assessed by Eq. 3 for different H structures in different bcc metals under arbitrary stress states.

### 3.4 Nanosized hydride particles in bcc metals

Our discussion above has demonstrated the anisotropic nature of volume expansion (or contraction) and stress responses of different H structures. However, the DFT investigation is limited to interstitial H and ordered H cluster structures with periodicity (i.e., infinite planar H clusters or hydrides, see Fig. 1). In order to expand the scope and extend the investigation to more practical situations, such as nanohydride particles in metal and H clustering around defects (e.g., dislocations), we resort to atomistic (MD and/or GCMC) simulations based on empirical EAM potentials. In our EAM based simulations, we focus on the W-H and Fe-H systems due to the broad applications of these two metals in H-containing environments. Benchmark calculations also show that these two EAM potentials can generally reproduce DFT results for H interstitials and clusters (see Appendix C for details), thus suitable for study H clustering at larger scale.

With the EAM potentials assessed, we proceeded to conduct large-scale MD simulations of more complex H structures, among which one particular category investigated are nanohydride particles in metals. Spheroid-shaped nanohydride particles in W and Fe were constructed, as shown in Fig. 6a. Note the hydride phase is taken to be the $HD_O^{MH}$ mono-hydride since it is always more stable than the $HD_T^{MH2}$ di-hydride phase (cf. Fig. 2). The nanohydride has its polar axis along [001], i.e., the four-fold symmetry axis of their formation volume tensor, with its geometry described by parameters polar radius ($h$) and equatorial radius ($r$). The zero-stress formation enthalpies of the nanohydrides with varying size (number of H in the hydride) and pole-to-equator ratio, i.e., $h/r$ ratio, were evaluated with the results shown in Figs. 6b and 6c. From the data, we note that the formation enthalpy generally decreases as the nanohydride size increases, quickly become energetically more favorable than TIS H. Meanwhile, for nanohydrides of a particular size, the ones with smaller $h/r$ ratios tend to be more stable, suggesting that nanohydride particles prefer thin platelet structures perpendicular to the [001] direction. This shape preference is well expected because the hydride has an anisotropic formation volume tensor with large $\Omega_{001}^H$ component, and consequently the particle will prefer not to grow along [001] direction to minimize distortion and thus strain energy.

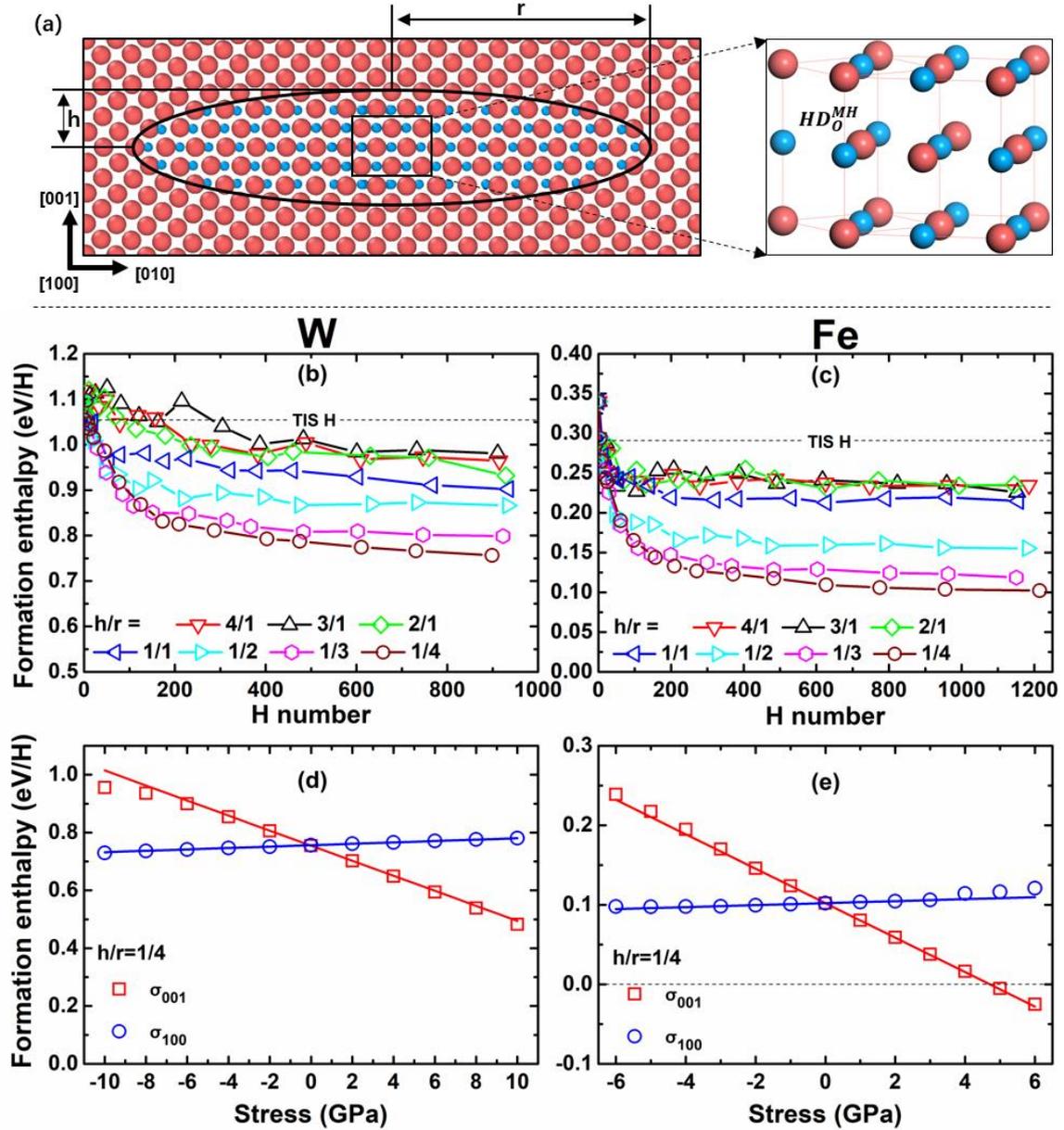

Figure 6. (a) Cross section of a spheroidal $HD_O^{MH}$ nanohydride particle in bcc metals, with its polar axis along [001]. Blue atoms are H while red ones are metal atoms. $r$ and $h$ respectively represent equatorial and polar radii of the spheroid. (b) and (c) respectively show zero-stress formation enthalpies of spheroid-shaped nanohydride particles in W and Fe, with varying size (i.e., H number) and pole-to-equator ratio (i.e., $h/r$). (d) and (e) respectively show the effect of uniaxial stresses along different directions (denoted by $\sigma_{001}$ and $\sigma_{100}$) on the formation enthalpy of representative nanohydride particles chose from (b) and (c) (largest ones for $h/r = 1/4$) in W and Fe. Symbols are results from MD simulations while lines indicate predictions based on Eq. 3.

In addition, we look into the effect of stress on those nanohydrides. Figs. 6d and 6e show the stress-dependent formation enthalpies for a representative nanohydride particle (largest ones in Figs. 6b and 6c for $h/r = 1/4$) when subjected to uniaxial stresses along different directions. We can see that the formation enthalpy shows highly anisotropic stress-dependence, being mostly affected by the [001] stress while not much by the [100] stress, which is not surprising considering the anisotropic formation volume tensor of the pristine $HD_O^{MH}$ hydride as previously discussed. Moreover, such stress-dependence of the nanohydride can also been well predicted by Eq. 3, indicating that Eq. 3 is applicable to arbitrary sized nanohydrides.

**3.5 Nanohydride formation around dislocations**

The preceding results have demonstrated that H clustering is energetically favorable and may be further assisted by anisotropic stresses in bcc metals. However, there is still a need to make the leap from energetics to thermodynamics. To this end, we performed a series of hybrid MD+GCMC simulations to assess the thermodynamic feasibility of H clustering at room temperature. We first performed simulations on the reference systems of stress-free pristine W and Fe bulk lattices at different H chemical potentials. Fig. 7 presents the relationship between equilibrium H concentration and H chemical potential obtained from hybrid MD+GCMC simulations, where the logarithm of H concentration shows an evident linear relationship to the H chemical potential. Consequently, the lattice H concentration $C_H$ can be fitted to Arrhenius type equations, i.e., $C_H = \exp(\frac{\mu+1.164}{k_B T})$ for W and $C_H = \exp(\frac{\mu+2.051}{k_B T})$ for Fe ($\mu$, $k_B$, $T$ respectively denotes H chemical potential, Boltzmann constant, and temperature). Note here H concentration is atomic concentration, namely the H/metal atomic ratio. With these Arrhenius relations determined, we can extrapolate H concentrations under any arbitrary H chemical potentials, even very low ones (which can be difficult to obtain directly due to the large simulation boxes required). In these simulations with pristine bulk W and Fe lattices, no H clustering was observed under all H concentrations explored, which is expected considering the nucleation barrier between random H solution and H clusters [35] and the limited time/step covered in simulations.

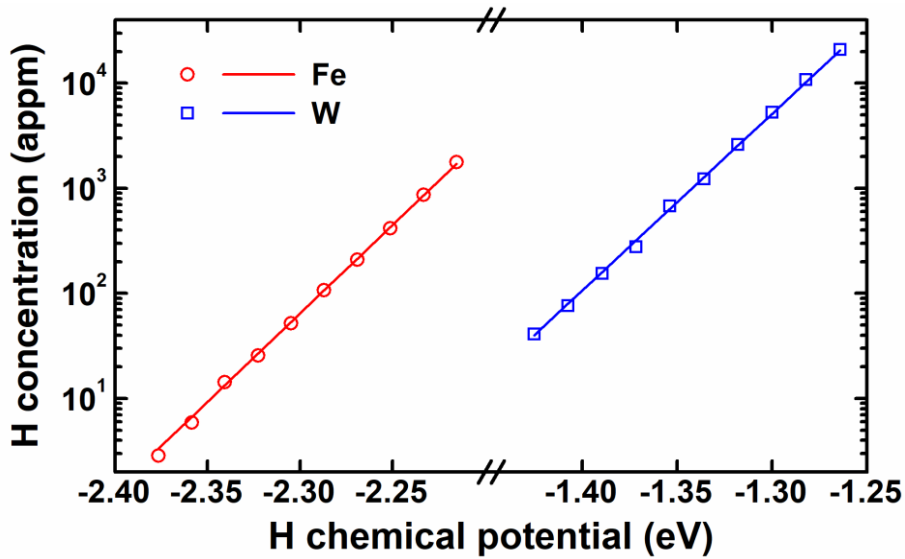

Figure 7. Relations between H concentration and H chemical potential in W and Fe at room temperature (i.e., T=300 K). Symbols are results from hybrid MD+GCMC simulations, while lines are fitted based on the Arrhenius equation.

The absence of H clustering in the pristine bulk W and Fe lattices indicates the need of additional driving force for H clustering to occur. As previously mentioned, one source of driving force may come from anisotropic stress field, which can be induced by either external loading (e.g., homogeneous uniaxial loading, see sections 3.2 and 3.4) or originate from defects. Dislocations are a prevailing group of defects in metals. In particular, edge dislocations, with dilatation-type stress fields, are expected to provide nucleation sites for H clustering and may subsequently trigger H blistering in bcc metals [7, 10, 48, 49, 72]. In this regard, we performed hybrid MD+GCMC simulations to examine H distribution and aggregation around edge dislocations in W and Fe. Here the edge dislocations of $(100)[001]$ and $(01\bar{1})[111]/2$ are selected as representatives, considering the recent experimental attentions paid to the former one [7] and the ubiquitous presence in bcc metals of the latter one.

Fig. 8 shows the MD+GCMC simulation results around $(100)[001]$ edge dislocations in W and Fe, at different H chemical potentials. For the case of W, clear H aggregation around the dislocation has been observed for all the H chemical potentials considered, with most H atoms forming platelet-like nanohydrides of an internal $HD_O^{MH}$ structure. These nanohydrides are located mostly on the tension side of the dislocation, clearly benefiting from the lattice dilation. We should note that in Figs. 8a and 8b with high H chemical potential (i.e., $\mu \geq -1.482$ eV or lattice H concentration $C_H \geq 4.5$ appm), the nanohydride formed would continue to grow without converging to a stable hydride size (even after $4 \times 10^7$ steps). This indicates that spontaneous growth of hydride would follow after the initial nucleation from the dislocation, leading to possible large-scale hydride formation. On the other hand, for low H chemical potential (i.e., $\mu \leq -1.601$ eV or lattice H concentration $C_H \leq 45$ appb ), the nanohydride around the dislocation would grow and then stabilize at a certain size, e.g., cases demonstrated in Figs. 8c and 8d. Nevertheless, even with an extremely low $C_H$ of 0.45 appb, there are still significant amount of H trapped around the dislocation (>150 H/nm), much higher than previous estimations based on isotropic elastic theory without considering H clustering [48, 49]. Such huge difference in number of trapped H will certainly affect H interaction with dislocations (and with other defects such as cracks or inclusions), necessitating further dedicated efforts to evaluate related influence on H damage mechanisms. Similar to W, for the case of Fe (c.f., Fig. 8e-h), we can see increasing degree of H accumulation at the $(100)[001]$ dislocation as H chemical potential (or $C_H$) rises. Nonetheless, nanohydride formation in Fe appears to demand much higher levels of $C_H$, with H distribution around the dislocation showing slightly different morphologies compared to those in W. The difference may be attributed to much smaller energy release for $HD_O^{MH}$ hydride formation with reference to lattice H (TIS H) (c.f., Fig. 2), and that Fe having much lower moduli than W, and thus a much weaker stress field to drive H clustering at the dislocation.

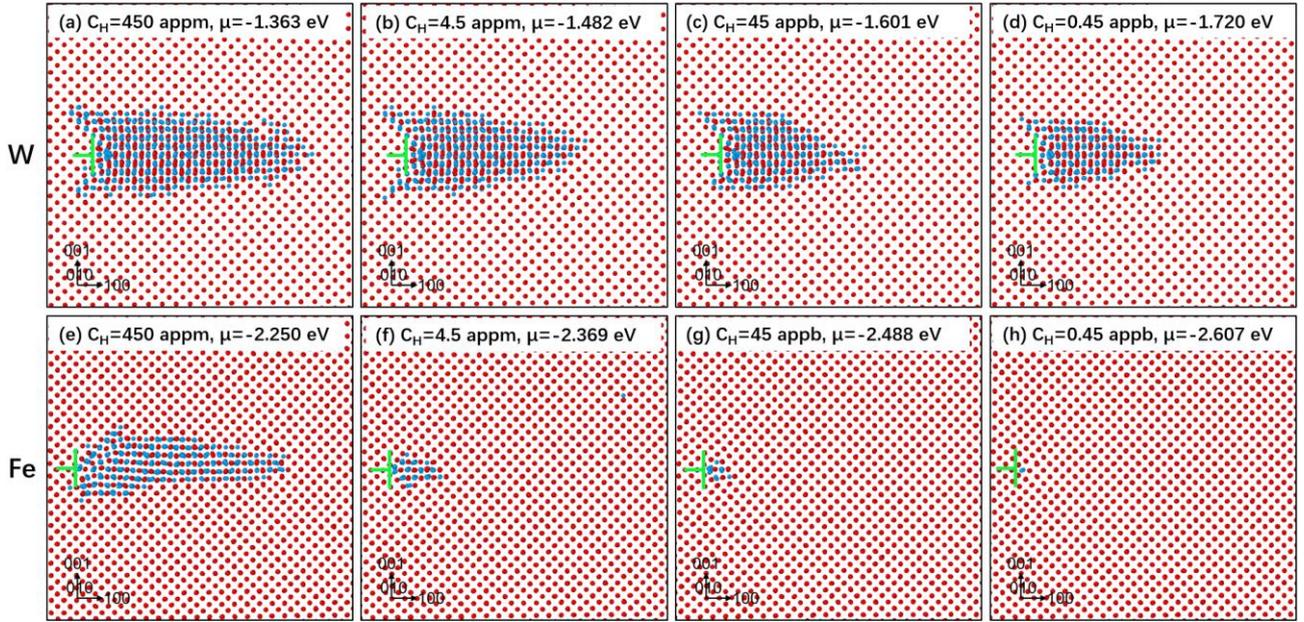

Figure 8. H distribution around $(100)[001]$ edge dislocations in (a-d) W and (e-h) Fe at 300 K. $\mu$ and $C_H$ denote H chemical potential and far field lattice H concentration respectively. Blue and red atoms indicate H and metal atoms respectively. Note that results in (a-b) and (e) are the configurations taken at $4 \times 10^7$ steps as the growth of nanohydride did not converge (indicative of possible spontaneous growth into large-scale hydride). Other configurations, i.e., c-d and f-h, are converged configurations.

Fig. 9 shows the results for the case of $(01\bar{1})[111]/2$ edge dislocation. A few similar observations of H aggregation and nanohydride formation can be made. Again, we can note preferential aggregation of hydrogen on the tension side of the dislocation, and that higher levels of $C_H$ are required in Fe than W for H clustering to occur. Similar spontaneous hydride growth was also indicated (hydride size does not converge even after $4 \times 10^7$ steps) in Figs. 9a, 9b, and 9e with high $C_H$. However, comparing the nanohydride sizes shown in Fig. 8 and Fig. 9, we note that at the same $C_H$ level, there is significantly more H aggregation/clustering around $(100)[001]$ than $(01\bar{1})[111]/2$. The difference is further demonstrated in Fig. 10, where we compare the size evolution of nanohydride as a function of far field H concentration for the two dislocations. Here the nanohydride size is represented by the number of H per unit length of dislocation. The profound difference in nanohydride size shown in Fig. 10 cannot be simply explained by the small disparity (~13.4%) in the burgers vector. Rather, it was attributed to the significantly smaller $\sigma_{001}$ or $\sigma_{010}$ components (normal stresses along [010] or [001] directions) from the <111>/2 dislocation compared to the <001> dislocation (see Figs. 11a-c). As previously discussed, the tensile component along one of the <001> directions is the key to facilitate H clustering. Therefore, nanohydride growth is expected to favor <001> type edge dislocations over <111>/2 type ones. Our results are in good agreement with recent experimental results [7] suggesting that <001> type edge dislocations play a more prominent role than <111>/2 ones in inducing H clustering and trigging H blistering.

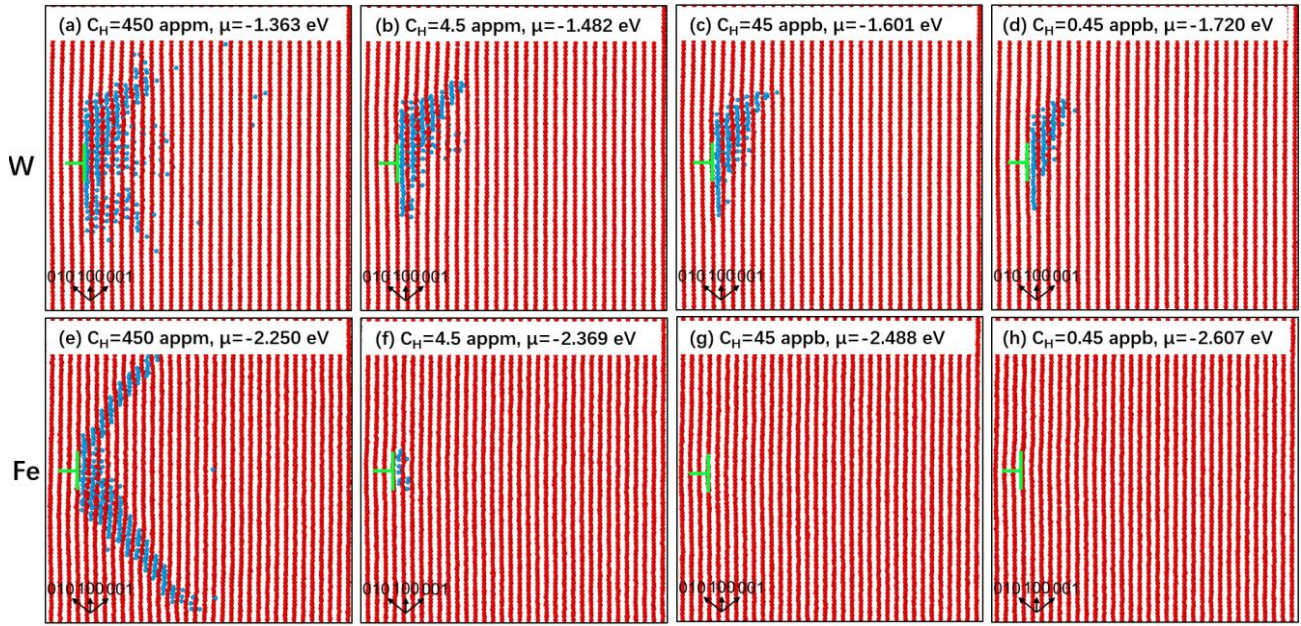

Figure 9. H distribution around $(01\bar{1})[111]/2$ edge dislocations in (a-d) W and (e-h) Fe at 300 K. $\mu$ and $C_H$ denote H chemical potential and far field lattice H concentration respectively. Blue and red atoms indicate H and metal atoms respectively. Note that results in (a-b) and (e) are the configurations taken at $4 \times 10^7$ steps as the growth of nanohydride did not converge (indicative of possible spontaneous growth into large-scale hydride). Other configurations, i.e., c-d and f-h, are converged configurations.

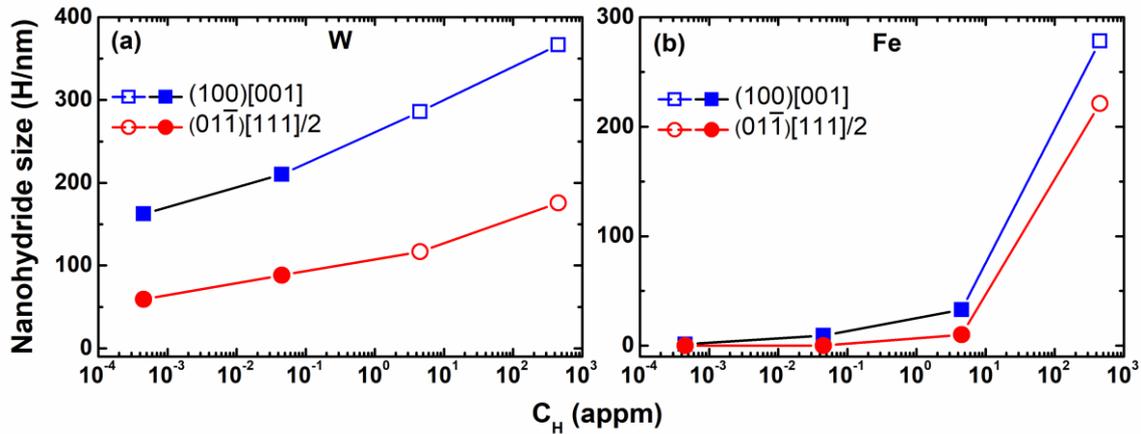

Figure 10. The resultant nanohydride size (represented by number of H per unit length of dislocation) around $(100)[001]$ and $(01\bar{1})[111]/2$ edge dislocations in (a) W and (b) Fe at 300 K. $C_H$ denote far field lattice H concentration. Note that solid symbols indicate cases where nanohydride growth was converged while hollow symbols indicate cases where nanohydride growth did not coverage at $4 \times 10^7$ steps (and possible spontaneous growth beyond into large-scale hydride), and in those cases the nanohydride size is taken as the one obtained at $4 \times 10^7$ steps.

Meanwhile, another significant distinction we may note comparing Fig. 8 and Fig. 9, is that nanohydrides formed around [001] dislocations extend mostly along the (001) half-plane where normal tensile stresses dominate with negligible shear stresses (see Fig. 11a). While around [111]/2 dislocations, the nanohydride first grows with H aggregation within the (01$\bar{1}$) slip plane, which at first sight seems to be a shear-dominating region in the default x/y/z coordinate system (see Fig. 11d). Nevertheless, in a lattice frame with basis vectors along [100]/[010]/[001] directions, the (01$\bar{1}$) slip plane is actually dominated by normal stresses resolved along <001> directions (see Figs. 11b-c), which favors H clustering. This actually suggests that caution is necessary when coming to interpreting the effect of stress on H clustering. Related to this, it is worth noting that there have been many previous studies showing that shear stress/strain could notably affect behavior of interstitial H [73-75] or nanohydrides [4, 37] in bcc metals, meanwhile those effects reported were found to show large variation, being highly orientation dependent [73-75], and diminish when the shear is applied along {001} planes [74]. Further examining the results shown in Fig. 9, we note that after a few layers of H accumulation along the (01$\bar{1}$) slip plane, further growth of the nanohydride would branch out to extend roughly along {012} planes. Such growth pattern was also reported in a recent MD study [37]. In addition, similar anisotropic stress enhanced H clustering will likely occur around other microstructural entities such as crack [4] or inclusion [69], necessitating further dedicated work to evaluate its impact on H induced damages in bcc metals.

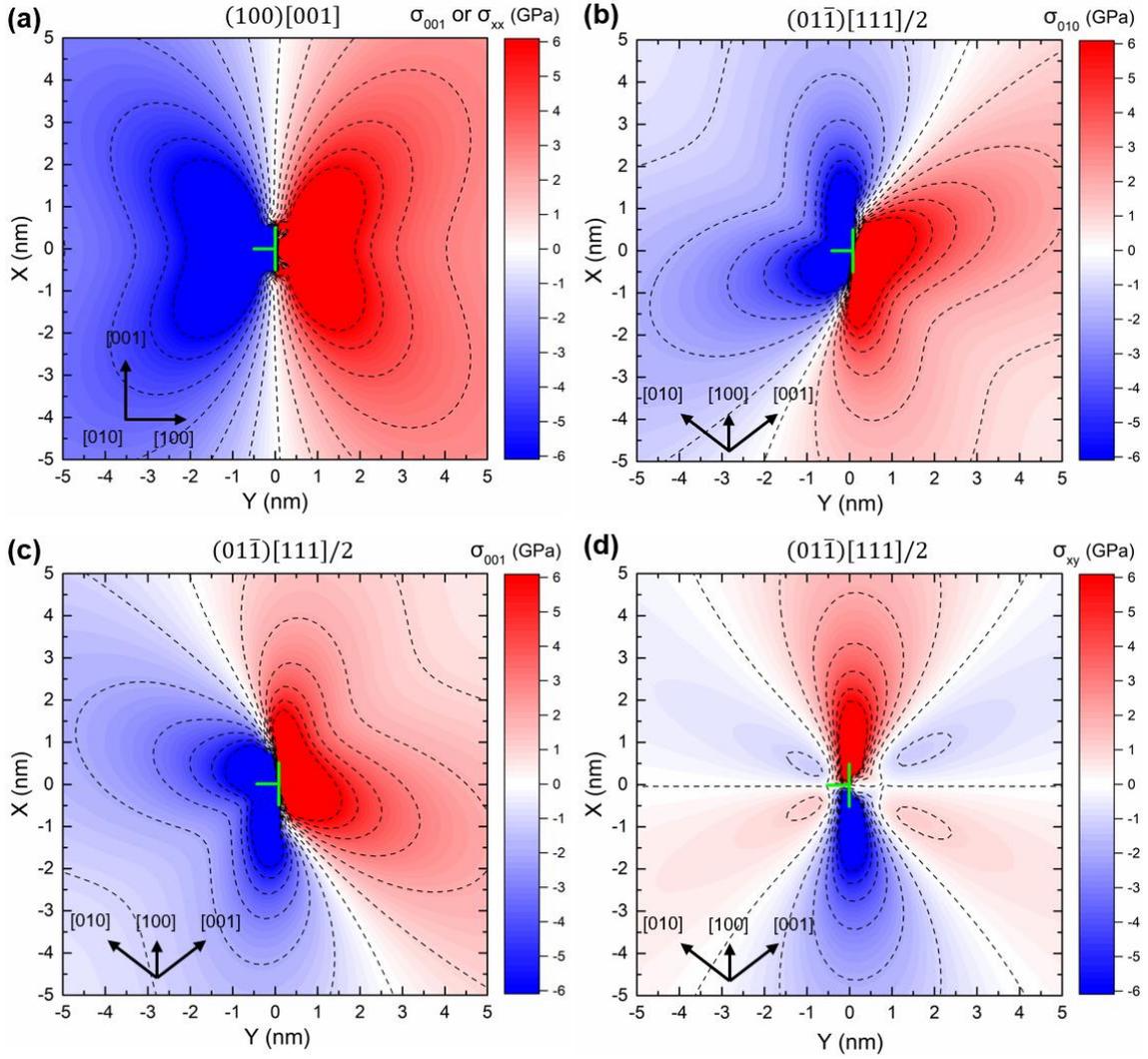

Figure 11. Elastic stress field around different edge dislocations in W from MD simulations at 0K, showing (a) $\sigma_{001}$ (i.e., $\sigma_{xx}$) stress component around a $(100)[001]$ edge dislocation, and (b) $\sigma_{010}$, (c) $\sigma_{001}$ and (d) $\sigma_{xy}$ stress components around an $(01\bar{1})[111]/2$ edge dislocation.

## 4. Conclusions

In this work, H clustering behaviors in non-hydride-forming bcc metals represented by W, Fe, Mo, and Cr, have been systematically investigated combining first-principles calculations, atomistic and grand canonical Monte Carlo simulations. Based on our results, the following conclusions can be made:

1. H exhibits much lower formation enthalpies in aggregated structures (i.e., ordered planar H clusters and hydrides) than its interstitial forms, indicative of H clustering being energetically favorable in bcc metals;

2. The effect of stress on H energetics in different structures was found to be highly anisotropic, particularly favoring tensile stress component along one of the <001> crystalline directions, yet insensitive to other stress components;

3. This anisotropic stress effect was elucidated to originate from the anisotropy in corresponding H formation volume tensor, and shown to be well predicted by the continuum linear elasticity model for any arbitrary stress state;

4. Nanohydride growth simulations at two typical dislocations in bcc metals, i.e., (100)[001] and $(01\bar{1})[111]/2$, showed that the dislocation stress field can induce nanohydride formation at rather low levels of H concentration or chemical potential, and that spontaneous nanohydride growth would occur when H concentration or chemical potential exceeds certain threshold, confirming the thermodynamic possibility of large-scale H clustering assisted by defects;

5. The nanohydride at dislocations was found to grow in the form of thin platelet structures that maximize one <001> tensile stress component, in accordance with anisotropic stress effect previously demonstrated. This explains the different sizes and morphologies of nanohydride formed at (100)[001] and $(01\bar{1})[111]/2$ dislocations. It also indicates that (100)[001] dislocations, with much larger <001> tensile components, would play a more important role in H clustering in bcc metals, in close agreement with recent experimental observations.

Our present study explicitly and quantitatively elucidated the anisotropic nature of stress effect on H energetics and H clustering behaviors, providing new insights critical towards understanding H-induced damages in metals.

**Acknowledgement**

This work was financially supported by Natural Sciences and Engineering Research Council of Canada (NSERC) Discovery Grants Program (grant #: NSERC RGPIN-2017-05187), the McGill William Dawson Scholar Award. X.S. Kong and C.S. Liu acknowledge the financial support from the National Key R&D Program of China (grant no. 2018YFE0308102) and the National Natural Science Foundation of China (Nos.:11735015, 51771185). J. Hou and J. Song acknowledge Compute Canada and the Supercomputer Consortium Laval UQAM McGill and Eastern Quebec for providing computing resources.

**Appendix A. Convergence tests for super-cell size**

Benchmark calculations with different super-cell size were performed to ensure there is no size dependence in our results. These calculations were carried out using EAM based MD simulations considering the size dependence are generally associated with classical elastic interactions. Using W-H system as a representative, we calculated formation enthalpies and volume tensors for interstitial H and (001) planar H clusters, with varied super-cell size respectively along all three and [001] directions (ordered hydrides are periodic along three dimensions thus are not affected by super-cell size). As shown in Fig. A1, the supersize used in the main text is sufficiently converged, with no notable change in both formation enthalpy and volume tensor upon further increase in supercell size. Note convergence tests for plane wave cutoff and k-point density were done in our previous work [27], thus are not shown here.

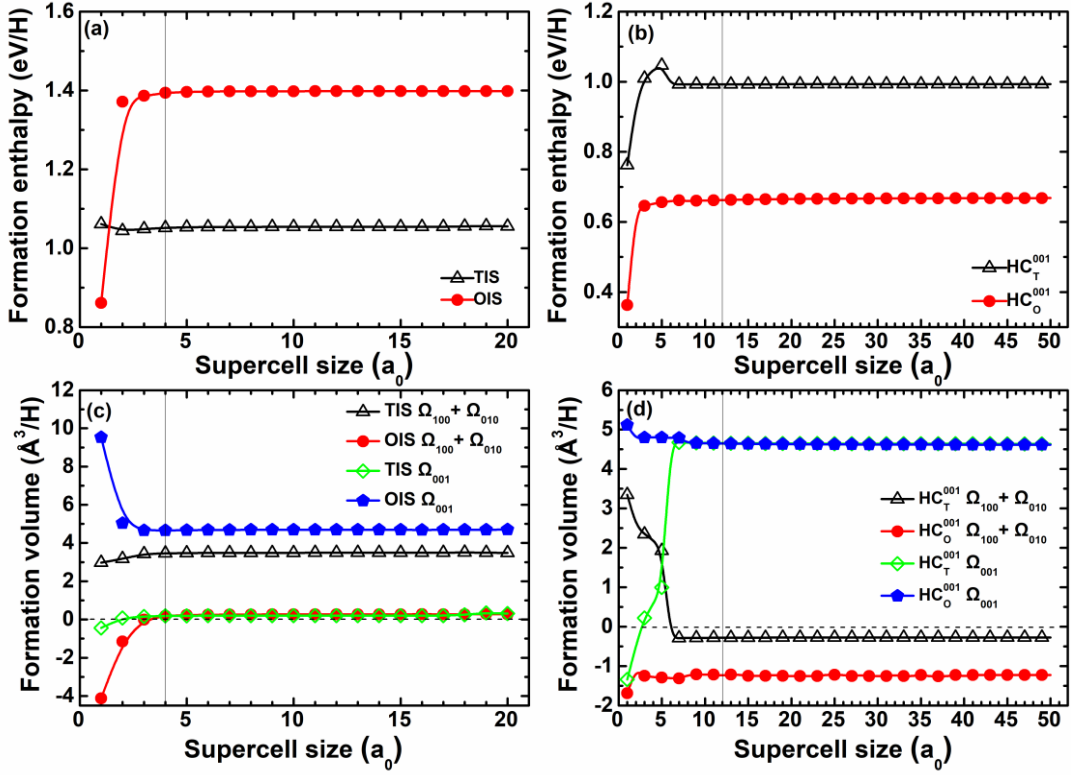

Figure A1. Formation enthalpies for interstitial H (a) and planar H cluster (b) in W, and related formation volume tensor components (c-d), as functions of super-cell size. Results were calculated using EAM potential based MD simulations. Vertical lines highlights super-cell sizes used in the main text.

**Appendix B. Different methods for predicting H formation enthalpy under anisotropic stress or strain**

In the main text, we predict H formation enthalpy under an external stress $\widetilde{\sigma}^{ext}$ by Eq. 3. As demonstrated in previous works [70, 71], equivalent predictions can also be made by:

$$H_f^H(\varepsilon^{ext}) = H_f^H(0) - \widetilde{\varepsilon}^{ext} : \widetilde{P}^H(0), \quad (B1)$$

where $\widetilde{\varepsilon}^{ext}$ is the external strain induced by $\widetilde{\sigma}^{ext}$, and $\widetilde{P}^H(0)$ is the elastic dipole tensor of H under zero strain. In linear elastic region, $\widetilde{P}^H$ and $\widetilde{\Omega}^H$ are correlated by:

$$P_{ij}^H = -V_{bulk}\sigma_{ij}^H = V_{bulk}C_{ijkl}\varepsilon_{kl}^H = C_{ijkl}\Omega_{kl}^H, \quad (B2)$$

where $\sigma_{ij}^H$ is the tensor component of H induced stress in a super-cell with fixed size and shape, and $C_{ijkl}$ denotes the stiffness tensor of the metal. Fig. B1 shows benchmark results comparing Eq. 3 and Eq. B1. Apparently, Eq. 3 and Eq. B1 provide almost identical predictions and agree excellently with EAM results, with only a small discrepancy due to non-linearity between $\tilde{\varepsilon}^{ext}$ and $\tilde{\sigma}^{ext}$ at large stress region.

Despite the two methods being equivalent in linear elastic regions, Eq. 3 is more convenient with straightforward insights in evaluating anisotropic behavior. This is because different components of $\widetilde{P}^H$ are coupled due to the Poisson effect (i.e., $C_{12} \neq 0$ in Eq. B2), therefore $\widetilde{P}^H$ does not reflect anisotropy as directly as $\widetilde{\Omega}^H$ does. This can be seen in Table B1 where $\Omega^H_{100}$ and $\Omega^H_{010}$ are almost zero for OIS H in W, indicating OIS H is not sensitive to stresses applied on these two directions according to Eq. 3. However, this insensitivity is not obvious from Eq. B1, considering $P^H_{100}$ and $P^H_{010}$ are quite significant for OIS H.

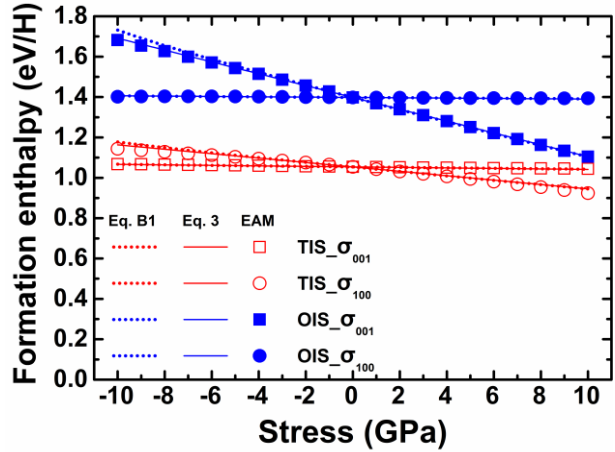

Figure B1. Formation enthalpies for H at TIS and OIS in W under uniaxial stresses along [001] and [100] directions (denoted by $\sigma_{001}$ and $\sigma_{100}$). Symbols are simulation results based on EAM potential. Solid lines are theoretical predictions from Eq. 3. Doted lines are theoretical predictions from Eq. B1.

Table B1. Tensor component of formation volume $\Omega^H_{ij}$ and elastic dipole $P^H_{ij}$ for OIS H in W, calculated using EAM potential under zero-stress condition.

|  | 100 | 010 | 001 |
|---|---|---|---|
| $\Omega^H_{ij}$ (Å³) | 0.13 | 0.13 | 4.69 |
| $P^H_{ij}$ (eV) | 6.72 | 6.72 | 16.29 |

**Appendix C. Comparison between DFT and EAM results**

One critical aspect of simulations based on EAM potentials is the accuracy and validity of the potential in describing interatomic interactions. In this regard, we first conducted a series of benchmark tests to compare the results from MD simulations with those from DFT calculations. Fig.

C1 compares the H formation enthalpy $H_f^H(0)$ and volume tensor $\widetilde{\Omega}^H(0)$ for H in different structures in W and Fe under stress-free condition, obtained from MD simulations and DFT calculations. For the W-H system, MD and DFT give similar formation enthalpies for TIS H and OIS H, but show difference on the formation volume tensors. Specifically, for TIS H, the W-H EAM potential underestimates $\Omega_{001}^H$ (0.2 Å$^3$ against 1.05 Å$^3$ by DFT) and overestimates $\Omega_{100}^H$ (1.75 Å$^3$ against 1.12 Å$^3$ by DFT); while for OIS H, both $\Omega_{001}^H$ and $\Omega_{100}^H$ are overestimated. Nevertheless, the W-H EAM potential manages to capture general trends of DFT results, as shown in Fig. C1a and C1c, thus is expected to remain capable of reproducing general features of H clustering. For the Fe-H system, the predictions by the EAM potential yields very good agreement with DFT data, therefore deemed to offer good description of H clustering behaviors.

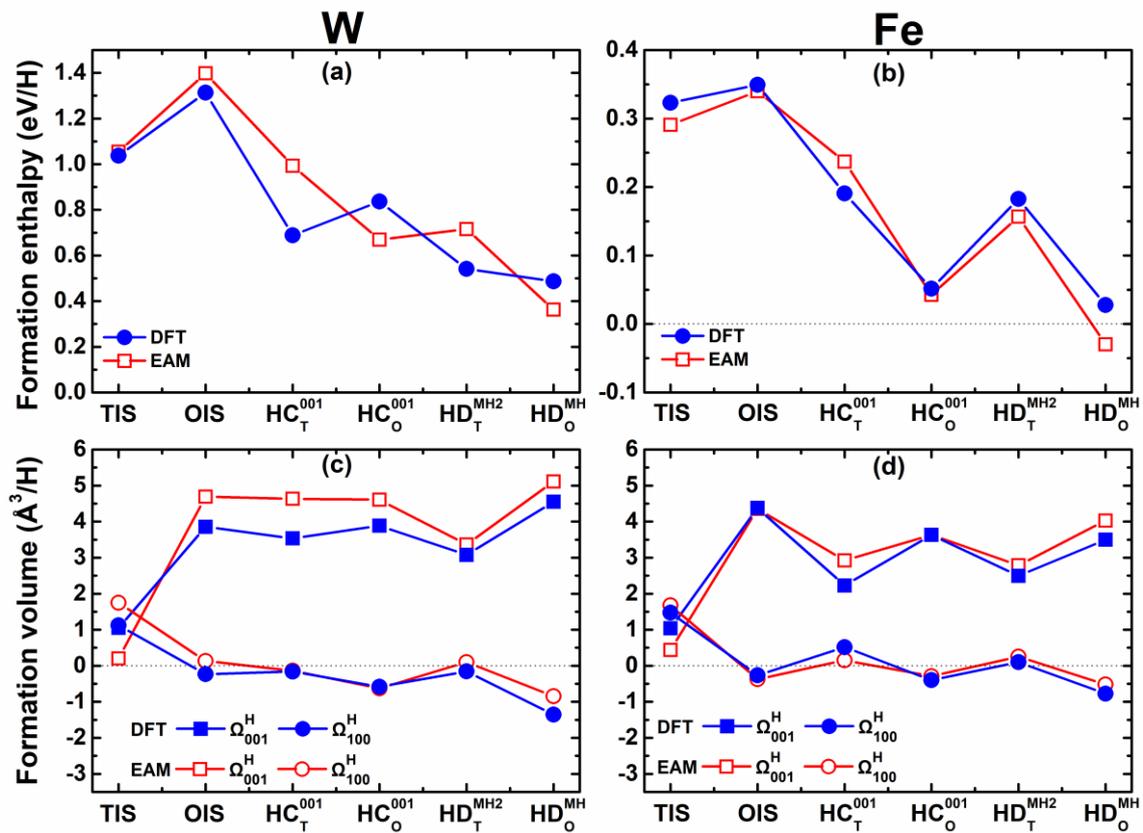

Figure C1. Comparison between DFT and EAM potential predictions of the H formation enthalpy $H_f^H(0)$ for interstitial H, planar H clusters and hydride phases in (a) W and (b) Fe, and the associated formation volume tensor components for (c) W and (d) Fe.